# Novel software for continuous wavelet analysis enable EEG real-time analysis on portable computers

Shoichiro Nakanishi

Department of Neuropsychiatry, Graduate School of Medical Sciences, Kyushu University, Fukuoka, Japan

## Abstract

Continuous Wavelet Transform (CWT) is frequently used for waveform analysis. For example, in the field of neuroscience research, CWT is performed to analyze electroencephalograms (EEG) and calculate the index of brain activity. Recent advancements in computer technology, such as general-purpose computing on Graphics Processing Units (GPGPU), have enabled the application of CWT to real-time waveform analysis. However, the computational complexity of CWT is large, and it is challenging to employ CWT as a real-time analysis method, such as in brain-machine interfaces (BMI), which require small size and cost. Therefore, a fast calculation method suitable for small and lightweight computers is desired. In this study, Python-based software for the CWT was developed and tested on portable computers. Using this software, real-time analysis of 64-electrode EEG data based on CWT was simulated and demonstrated adequate speed for the real-time analysis. Furthermore, it exhibited flexibility in performing CWT with various parameters. This software can contribute to the development of compact and lightweight BMI devices. Since CWT is a mathematical method, it may be used as a tool for other purposes.

## Introduction

There are various methods for analyzing brain waves. For example, averaging (Pfurtscheller and Silva 1999), discrete Fourier transform, and entropy are used to calculate the indices. However, these methods cannot analyze both time and frequency axes. Brain waves are classified according to their function and frequency into delta, theta, alpha, beta, and gamma waves, and therefore, the frequency axis is important. Wavelet transform is a method for analyzing waveforms that can analyze not only the frequency axis but also the time axis. However, multiresolution analysis (MRA) or discrete wavelet transform (DWT) (Daubechies 1992) are kind of the wavelet transform, and the resolution of the frequency axis is rough. Furthermore, these methods can only analyze wave form of which length is multiply of power of 2. The continuous wavelet transform (CWT) is a kind of wavelet transform that can analyze the frequency axis in detail and is used in EEG and MEG studies to analyze brain activity (Tallon-Baudry and Bertrand 1999, Cohen 2014).

Recently, BMI has been actively studied, and considerable efforts have been made to design better real-time decoding algorithms (Lebedev and Nicolelis 2017). Fast Fourier transform (FFT) (Cooley and Tukey 1965; Rader 1968), which is an algorithm for calculating the discrete Fourier transform fast, is performed as a method for EEG analysis without a time axis. CWT is one of the methods that can be used for BMI (Kant et al. 2020), and can be performed on every possible scale of frequency. However, CWT requires $O(n^2)$ computational order. The speed of CWT can be accelerated using FFT algorithm, whose computational order is $O(n \log n)$, and cyclic convolution theorem. Even though CWT can be accelerated, it requires big amount of computation (Su and Xu 2010) and this feature is a major limitation for CWT as a method of BMI (Wang, Veluvolu, and Lee 2013).

In recent years, parallel computation techniques have been developed, which may provide a solution to the problem of computational complexity. For example, multi-threading and multi-processing are major parallel computation techniques for central processing units (CPU). Multi-threading (Marr 2002) is a technique that uses multiple threads in one processor, and multi-processing is a technique that uses multiple CPU cores in one task, which makes use of the performance of multicore CPUs. However, using all CPU cores with such techniques, shared memory, CPU cache, or inter-process communication can limit the performance of the software. On the other hand, the graphics processing unit (GPU), which is a part of the computer, made it possible to accelerate parallel computation significantly, and this technique is called "general-purpose computing on graphics processing units" (GPGPU). Using this technique, a detailed real-time EEG analysis of the whole brain by CWT became possible (Efitorov et al. 2018) (Chen et al. 2017) (Deng et al. 2012). In the case of GPGPU, transferring data between the main memory and GPU memory can limit performance.

Size, cost, and complexity are the problems associated with BMI (McCrimmon et al. 2016). Regardless of its high computational complexity, BMI systems based on CWT must be small, flexible, and inexpensive. Large computers are frequently used in real-time wavelet analysis, however, I believe that, small computers such as laptop computers, mobile phones, or embedded systems perform real-time analysis. I believe that, it is important to apply the knowledge obtained from research on large devices to the small device for clinical or daily use without modification. Furthermore, multiple CPU architectures should be supported because x86 or x64 architectures are widely employed in large computer systems, and arm architectures are widely employed in small computer systems. To solve these problems, efficient software (ninwavelets) (https://github.com/uesseu/ninwavelets) was developed using Python (Rossum and Boer 1991) and C language, which can work on x86, x64, arm architecture CPU, and nvidia GPU optionally. CWT was performed on an x86 architecture laptop computer and a small single-board computer, which is an ARM-based architecture. To confirm the flexibility and robust computation of the CWT, various parameters were measured, and the speed of computation was measured. To apply research in clinical practice, real-time and lightweight software should return the same results as detailed software. The results of the ninwavelets were compared to those of MNE-python (Gramfort et al. 2013), which is widely used in EEG/MEG studies.

## MATERIALS AND METHODS

### Fourier transformed formula

The calculation of CWT involves convolution. The computational order of convolution is $O(N^2)$ and difficult to use in real-time analysis. There is a method to reduce the computational cost using FFT based on the cyclic convolution theorem. It is known that the result of cyclic convolution is equal to the inverse Fast Fourier transform (iFFT) of the product of FFT results. Computational cost of Fast Fourier transform is $O(NlogN)$, and this method can be used for real-time analysis. In the case of CWT, both the wavelet and acquired data are first transformed by FFT, and then their products are transformed by iFFT.

$$\mathcal{F}^{-1}(\mathcal{F}(f(t)) * \mathcal{F}(\Psi(t)))$$

Where $\mathcal{F}$ is FFT, $\mathcal{F}^{-1}$ is iFFT, $\Psi$ is wavelet function, $t$ is time and $f$ is waveform function. Cyclic convolution approximates simple convolution, except around the edge of the wave. This method makes CWT faster than simple convolution, but needs to perform one iFFT and two FFT. Using the Fourier transformed formula of wavelets, one iFFT and one FFT are required.

$$\mathcal{F}^{-1}(\mathcal{F}(f(t)) * \widehat{\Psi}(w))$$

Where $\widehat{\Psi}$ is Frourier transformed formula of wavelet and $w$ is frequency. Without FFT, $\widehat{\Psi}(w)$ can be calculated directly from formula. For example, Fourier transformed Morlet wavelet is given as follows.

$$\widehat{\Psi}_{\sigma \text{Morl}}(w) = \mathcal{F}(\Psi_{\sigma \text{Morlet}}(t)) = c_\sigma \pi^{-\frac{1}{4}}(e^{-\frac{1}{2}(\sigma-\omega)^2} - k_\sigma e^{-\frac{1}{2}w^2})$$

$\sigma$ is a parameter of Morlet wavelet. $k$ and $c$ are constants calculated from $\sigma$. If $k$ is zero, it is called Gabor wavelet, which is simple and easy to calcurate. Gabor wavelet approximates Morlet wavelet when $\sigma$ is large. Therefore, in some cases, the Gabor wavelet can be a substitute for the Morlet wavelet. The formula of Gabor wavelet is given as follows.

$$\widehat{\Psi}_{\sigma \text{Gab}}(w) = \mathcal{F}(\Psi_{\sigma \text{Gabor}})(t) = c_\sigma \pi^{-\frac{1}{4}}(e^{-\frac{1}{2}(\sigma-\omega)^2})$$

There are wavelets which is originally defined as Fourier transformed formula, such as the Generalized Morse wavelet(GMW) (Lilly and Olhede 2012). The Fourier-transformed GMW is given as follows.

$$\widehat{\Psi}(w)_{\text{GMW}} = sign(\omega)\alpha_{\beta\gamma}\omega^\beta e^{-\omega^\gamma}$$

$\beta$ and $\gamma$ are parameters and $\alpha$ can be calculated from $\beta$ and $\gamma$. In particular, there are some advantages to use the Fourier transformed formula in the case of GMW. Such wavelets are originally Fourier transformed, and calculating waveforms of such wavelets causes rounding errors, edge problems of wavelets, and long times to spend. Using a Fourier transformed formula may be better in the case of Morlet wavelets too, because one FFT can be skipped when using such formulas and the solution of edge problem is not arbitrary. For these reasons, Fourier-transformed formulas were employed.

## Fast Fourier transform

CWT can be performed by the convolution theorem, and thus the speed of the CWT depends on the FFT algorithm. FFT can be performed when the length of wave is a multiple of numbers by variations of Cooley–Tukey FFT algorithm (Cooley and Tukey 1965). Furthermore, Rader's FFT algorithm (Rader 1968) can perform FFT on waves whose length is a prime number. This shows that FFT is a flexible method. Cooley-Tukey FFT algorithm is very fast and can be used as a method for rapid cyclic convolution. It is possible to transform the padded waveform to compute the cyclic convolution faster when the length of waveform is not multiply of small numbers. Paradoxically, padding wave to power of 2, which is often used length, may require the allocation of a large memory space and this might take a long time. Since it is known that varietes of Cooley-Tukey FFT algorithms do not require power of two wavelengths, waves are padded to multiply of small numbers. On the other hand, MNE-python pads wave to power of two. The small numbers are 2, 3, 5 and 7, because they are hard coded in low level FFT library fftpack (https://www.netlib.org/fftpack/)(Swarztrauber 1982), and this is used in numpy (Harris 2020) (https://numpy.org/), a widely used Python package. The wavelength to pad was set to be the smallest multiply of small numbers after the length of wave to minimize the time to allocate memory. Since searching such numbers was slow on pure python, the function to search the numbers was written in C language(https://www.bell-labs.com/usr/dmr/www/chist.html), which is fast low-level language, and embeded into ninwavelets to pad automatically if option is set.

## Optimizing performance

However python (https://www.python.org/) is a frequently used computer language in scientific computation, it is not as fast as low level computer languages, which needs compilers like C or C++(https://isocpp.org/). Using packages which is written in low level languages like numpy is the way for python to perform high performance scientific computation. Cupy, which is a numpy like GPGPU package of Python, and developed fast CWT software. In recent years, GPGPU has made calculations extremely fast in some cases. However, GPGPU is not fast in all cases, since transferring data from main memory to GPU memory takes a long time. It is important to reduce transfer data between GPU memory and main memory for performance. In addition, it is also important for parallel-computing to share memory between threads or processes. First, Fourier transformed wavelets were generated in GPU memory. Then, EEG/MEG data were transferred to GPU memory and

CWT was performed. Finally, the results of CWT were transferred to main memory. To avoid transferring data between main memory and GPU memory, ninwavelets does not actively perform calculations by the CPU when they are already on GPU memory. Numpy and multi-threading, which is a technique for high-performance computation on multi-core CPU. It is known that, in the case of script languages such as Python or Ruby (*https://www.ruby-lang.org*), Global Interpreter Locking (GIL) limits multi-core CPU feature of multi-threading. Multi-processing, which is a technique that requires more resources than multi-threading, is frequently used when parallel computing is required in such languages. Furthermore, multi-processing requires interprocess communication, which makes calculations extremely slow in some cases. Numpy, which is written in low-level languages and can perform calculations without GIL, and high-performance parallelism based on multi-threading can be performed in Python.

## Devices

The wave analysis performance was tested using a laptop computer (CPU: Intel(R) Core(TM) i7-13700H 2.90 GHz, GPU: RTX 4060 Laptop CUDA cores 3072units, OS: Linux 6.9.7, GPU driver: nvidia-550.90.07-2). Because Core i7-13700H is an x64 architecture multi-core processor with 6 performance cores and 8 efficient cores, CWT was tested only on performance cores for stability, and clock frequency of CPU was set to 2.9 GHz. To confirm performance of parallel computing without GPU, CWT using single-thread, multi-process and multi-thread were tested. When CWTs were performed by multi-process or multi-thread procedure, EEG channels were divided to 6 groups and CWT was performed using 6 threads. To test performance on embedded device, Raspberry Pi 5 (*https://www.raspberrypi.org/*), an inexpensive ARM architecture single-board computer (SBC) sold as educational device was also tested. Because ninwavelet does not support GPU of Raspberry Pi 5 for GPGPU, only CPU performance of Raspberry Pi 5 was measured. 4 threads were used for test of CWT using Raspberry Pi 5 which has 4 core CPU. The other conditions were the same as those for the laptop computer. To measure the speed of making wavelets, time spent by making wavelets without performing CWT was also measured.

## Precision

To confirm precision of ninwavelets, the results of wavelet transform of EEG were plotted(Figure 1). Random 1 second wave with 1000 Hz sampling frequency was transformed by Morlet wavelet($\sigma = 7.0$), and power value was plotted(Fig.4). GMW($\beta =$

$17.5, \gamma = 3$) based CWT was also performed by ninwavelets and parameters of GMW were adjusted so that waveform was similar to that of Morlet. MNE-python version of GMW was not calculated because current version of MNE-python (1.7.0) has no function to perform CWT based on GMW. To evaluate precision, power value of 30~90 Hz frequencies with increments of 1 Hz was calculated, and 0.2~0.8 sec value was extracted. Real part, imaginary part and phase of the CWT results were extracted. Then, the result of CWT of ninwavelets was subtracted from that of MNE-python and 600 time points and 60 frequency (36000 time-frequency points) bands were calculated. The absolute value of difference between two results of software by standard deviation of result of MNE-python was calculated and if it was larger than $1 \times 10^{-5}$ SD, it was defined as an error. The calculation was repeated 1000 times and mean of the results was obtained.

## Comparison between devices and software

The speeds of devices, methods of parallel computing and software were compared. To measure speed, EEG data of which sampling frequency is 1000 Hz was generated and programmed to perform CWT for 30~90 Hz 10 times. For stability, the test was performed 50 times and the mean of the 50 values was calculated. MNE-python and ninwavelets are used. Multi-processing, multi-threading, single thread and GPGPU computation speeds were measured.

### Wavelength and flexibility

To confirm the practical limitation of wavelength, time spent analyzing various lengths of waves by CWT was measured. EEG data whose number of samples are from 100 to 2500 points were produced and sampling frequency was set to 1000Hz to simulate waves from 0.1 to 2.5 seconds length. From 30 to 90Hz frequencies, which increments are 1Hz, were transformed by Morlet wavelet. CWT was performed 100 times for each wavelength. GPU and multi-threading computation was performed by laptop computer and multi-threading was performed by SBC. Time spent making Fourier transformed mother wavelets before CWT is performed was also measured. The frequency of 1 second mother wavelets is from 30 Hz to 90 Hz in 1 Hz increments and 1000 Hz sampling frequency.

# Result

Comparisons of Morlet wavelet between ninwavelets and MNE-python, is presented in Table 1. The results show that the difference between the results of two software is smaller than $1.0 \times 10^{-5}$ standard deviation of result of MNE-python about real and imaginary part. The phase part showed an error of 1.18 %. Figure 1 shows the results of Generalized morse wavelets of ninwavelets, Morlet wavelet of ninwavelets and MNE-python.

Time spent by CWT is shown in Figure 2(laptop computer) and Figure 3(SBC). In the case of laptop computer, ninwavelets was faster than MNE-python. GPU is faster than CPU however, in some cases, multi-threading is fast, too. Paradoxically, multi-processing was extremely slow under all conditions.

Various wavelength processing speeds are shown in Figure 4. GPU exhibited a faster speed than CPU. CPU showed unstable speed on both the laptop computer and SBC compared to GPU on the laptop computer. It did not show much high speed when the wavelength was 2 and was almost linear to the wavelength. The time spent to make Fourier transformed 1 second mother wavelets (sampling frequency = 2048 Hz) from 30 Hz to 90 Hz in 1 Hz increments is shown in Table 2.

Figure. 1

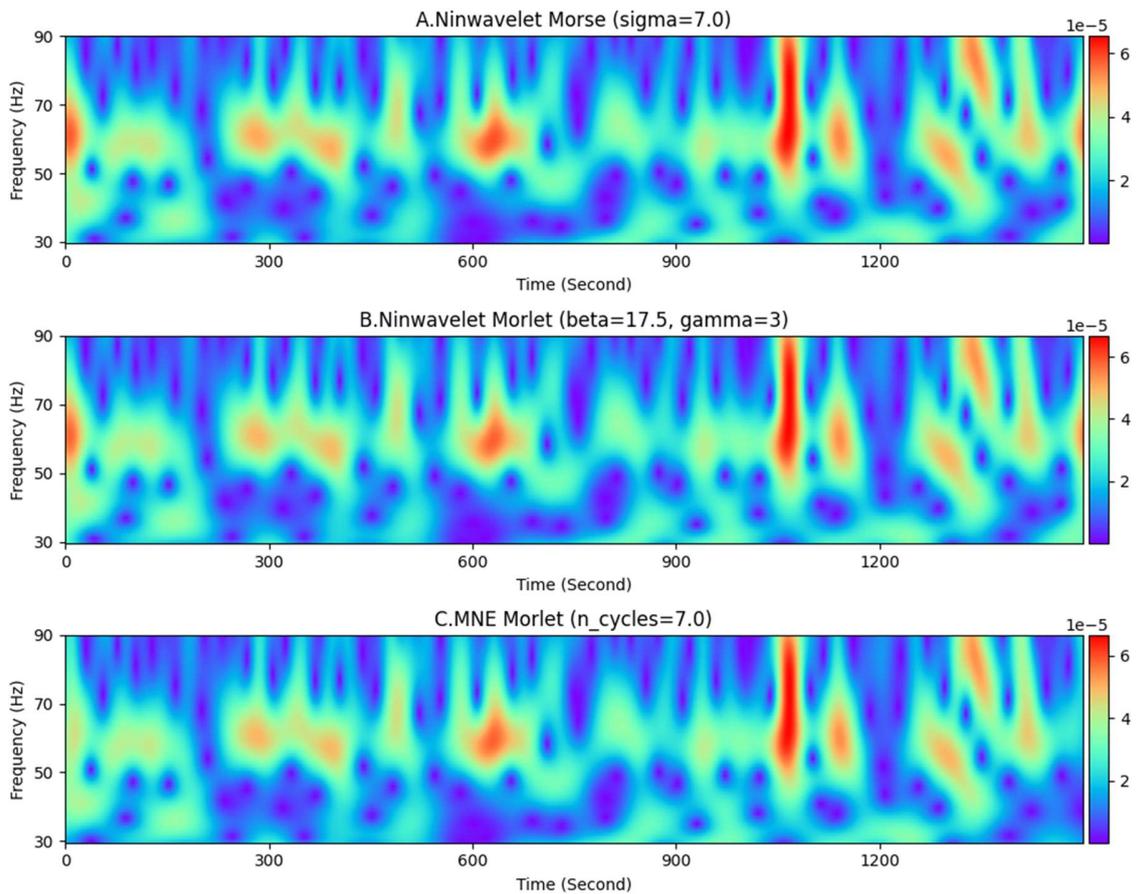

Plots of GMW and Morlet wavelets were calculated using ninwavelets and MNE-python. Σof Morlet wavelet was set to 7.0. 1 second wave was transformed from 30 Hz to 90 Hz in 1 Hz increments. The sampling frequency was set to 2048 Hz.

Table. 1

|  | Maximum difference | Mean difference | Error rate (times/test) |
|---|---|---|---|
| Real part | $7.00 \times 10^{-6}$ (SD) | $1.49 \times 10^{-6}$ (SD) | 0 |
| Imaginary part | $5.14 \times 10^{-6}$ (SD) | $1.07 \times 10^{-6}$ (SD) | 0 |
| Phase part | $3.17 \times 10^{-4}$ (Radian) | $1.56 \times 10^{-6}$ (Radian) | $1.18 \times 10^{-2}$ |

Difference between the CWT results calculated by MNE-python and ninwavelets. For the real and imaginary parts, the results were divided by the standard deviation of the result of MNE-python. In this table, 1 SD is the result of the standard deviation of MNE-python. Error points are points whose difference between the two software is bigger than $10^{-5}$ SD. The unit of the phase is radians.

Figure. 2

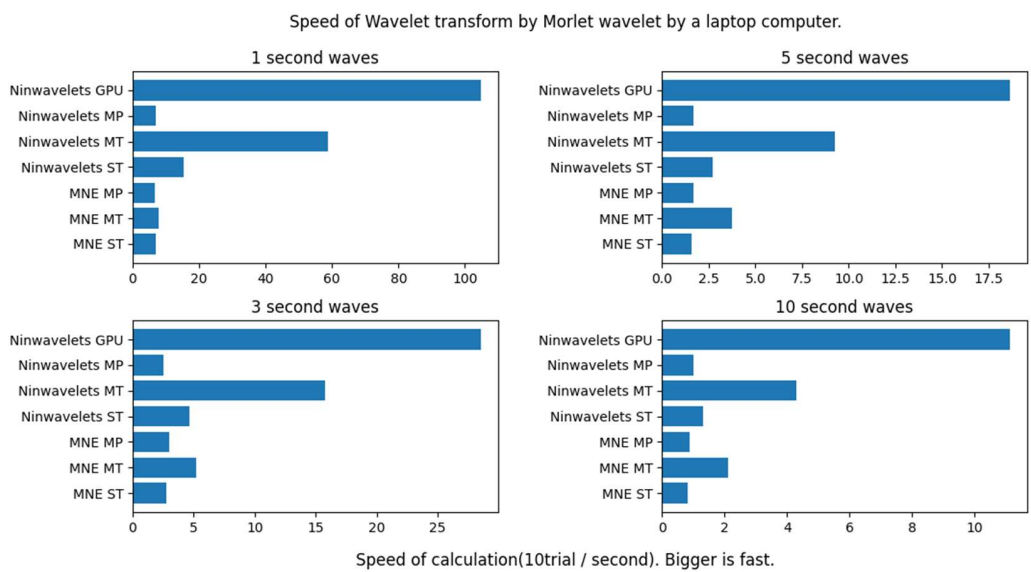

Calculation speed using ninwavelets and MNE-python. All the sensors of 64 channel EEG data were analyzed. The sampling frequency was 2048 Hz. A laptop computer with GPU is used. GPU means GPGPU. MP is multi-processing. MT is multi-threading. ST is single thread. 1, 3, 5, 10 second wave (sampling frequency= 2048 Hz) were processed by Morlet wavelet from 30 to 90 Hz for 1 Hz.

Figure. 3

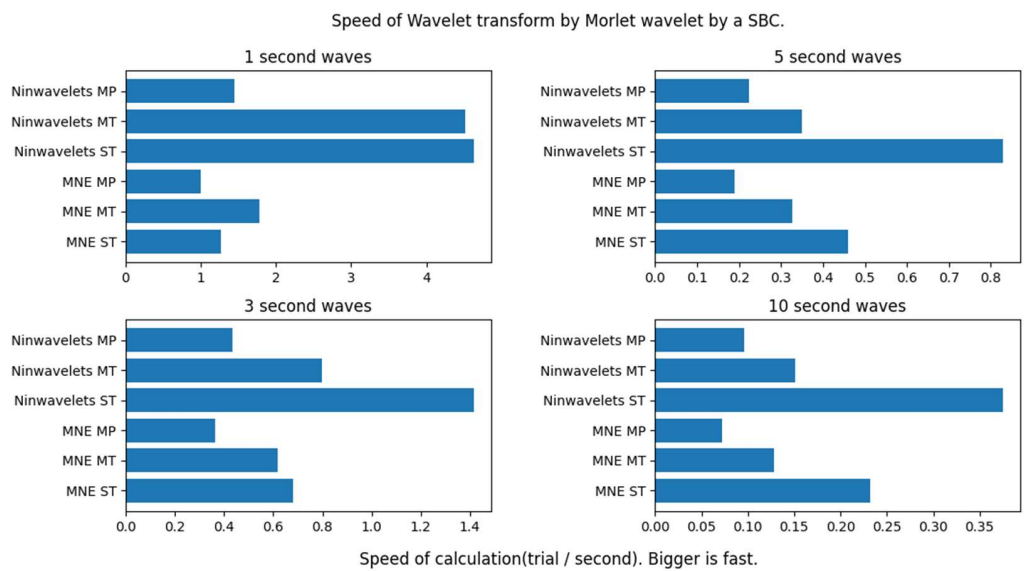

Speed of calculation by ninwavelets and MNE-python. Raspberrypi5 was used. All sensors of 64 channel EEG data was analyzed. The sampling frequency was 2048 Hz. MP is multi-processing. MT is multi-threading. ST is single thread. 1, 3, 5, 10 second wave (sampling frequency=2048Hz) was processed by Morlet wavelet from 30 to 90 Hz for 1 Hz.

Figure. 4

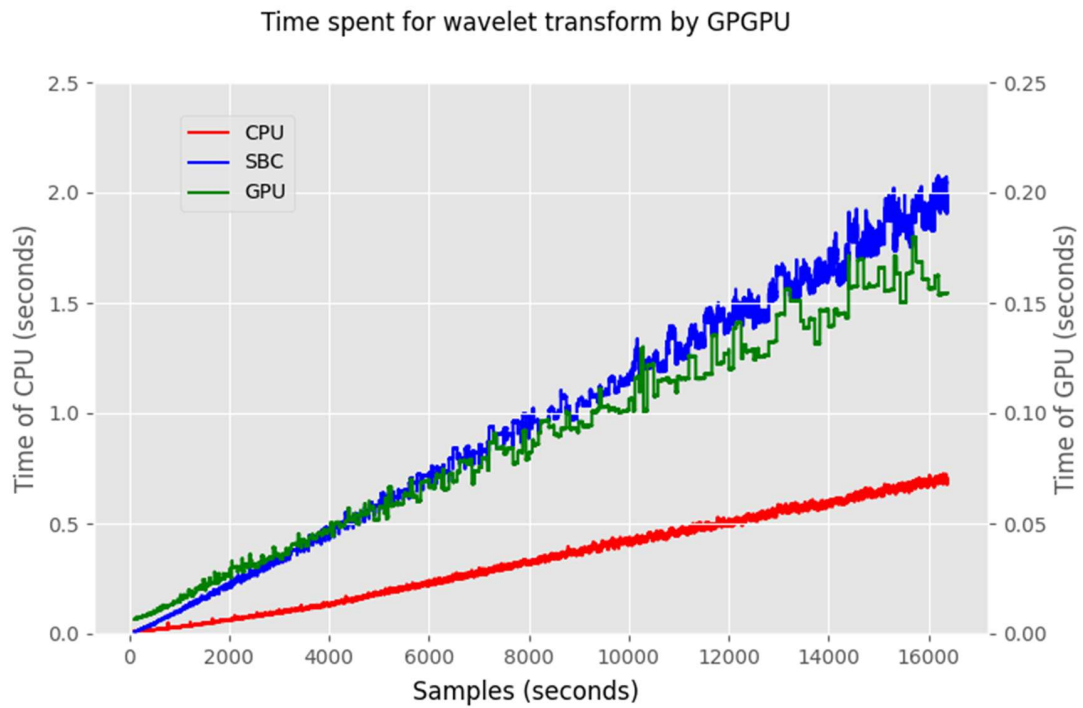

Time spent by ninwavelets CWT. CPU and GPU speeds were measured on a laptop computer, and CPU speed was measured on the SBC. When calculating using a laptop computer CPU, multi-threading was applied. When calculating using the SBC, single thread was used. Since GPU is fast, the axis of GPU speed is right, and axis of CPU speed is left. Each wave was processed 100 times.

Table. 2

| Wavelet | Device | Time(second) |
|---|---|---|
| Morse | Laptop GPU | 0.0227 |
| Morse | Laptop CPU | 0.00675 |
| Morse | SBC | 0.0112 |
| Morlet | Laptop GPU | 0.00432 |
| Morlet | Laptop CPU | 0.00970 |
| Morlet | SBC | 0.00497 |

Processing time to generate wavelets for 1 second 1000 Hz sampling frequency waves. The frequency ranges from 30 Hz to 90 Hz in 1 Hz increments.

# Discussion

## Performance

Figure 1 and Table 1 show the differences between ninwavelets and MNE-python. The difference between the real and imaginary parts of the wavelets is less than $10^{-5}$ SD and it seems to be sufficiently small. The phases are also similar to each other and mean of maximum difference is $3.17 \times 10^{-4}$ radian. This result seems to show adequate precision of ninwavelets for EEG processing. Furthermore, ninwavelets can use various wavelets, such as GMW, through inheritance ( object-oriented programming).

Figure 2 shows that ninwavelets has ability to process 10 seconds EEG data (64 channels, sanpling frequency=2048 Hz, 30 Hz to 90 Hz in 1 Hz increments) 10 times in 1 second and analyzed using a laptop computer GPU. Paradoxically, Figure 2 shows extremely low performance of multi-processing. The reason may be that inter-process communication, which is required in multi-processing of Python, took time to transfer huge results of CWT. However, multi-threading, which does not require inter-process communication, is faster than multi-processing. Figure 3 shows that the CWT speed of SBC and ninwavelets was faster than that of MNE-python. In the case of SBC, a single thread was faster than in the other methods. The architecture of computers is not the same as that of a laptop computer, and the reason for this cannot be simple. SBC could process 1 second EEG (64 channels, sampling frequency=2048 Hz) 4 times in 1 second. Even if the result of SBC is not much faster than that of MNE-python, such as GPGPU, real-time EEG analysis requires faster calculation and ninwavelets offers better performance. They were performed by the same program without modification, and these results may make real-time EEG analysis scalable.

Table 2 shows the time spent to generate Fourier-transformed 1 second mother wavelets (sampling frequency = 1000 Hz) from 30 Hz to 90 Hz in 1 Hz increments. Generating wavelets does not take much time, which suggests that it is possible to flexibly configure wavelets during real-time analysis.

## Length of wave

Since performance of FFT is affected by wavelength to analyze, this must cause limitation in parameters of CWT too. Figure 4 shows the result of CWT of waves whose length is a prime number (thought to be slow) and power of 2(thought to be fast), but there are no waves that are processed extremely fast or slow. We programmed to pad wave to multiply of small numbers and prime numbers are not problem. This shows that the power of 2 is not very important for the performance, and the process time seems to be linear to the wavelength. Computational order of FFT is $O(NlogN)$, and this means process time of FFT is not linear to wavelength, and so, the reason may be memory transfer and allocation speed. This suggests that, in the case of ninwavelets and computational complexity on the scale of EEG processing, users need not strictly consider the speed of FFT, and it is easy to develop real-time EEG processing systems.

## Advantages of CWT on portable computers

CWT is used widely in off-line analysis of EEG. If it is used as a method of BMI, knowledge of off-line EEG studies can be applied directly, and this may accelerate BMI studies. The results suggest that CWT-based BMI will be available on popular portable computers or embedded devices.

## Other usages

CWT is a mathematical method not only for EEG analysis. For example, analyzing drugs (Ertekin 2025), biomass of potato (Ruiz-Guzman 2025), displacement of a bridge (Nie 2025) and metro stray (Dong 2023). Ninwavelets is a simple mathmatical software for wavelet transform and can be used for signal processing in various fields.

## Conclusion

CWT based real-time analysis was available by previous studies. In this study, performance and flexibility of CWT on portable computers was tested. The result of CWT is a kind of feature value, and so this method can be combined with other methods. However, this study focused not on the entire BMI system but on basic software for analysis of wave form. Further development of BMI is desired.

## Acknowledgments

I am grateful to Dr. Shunsuke Tamura, Dr. Shogo Hirano, Dr. Takefumi Ueno, and Dr. Yhoji Hirano for useful discussions. I wish to thank Yuta Inoue for reading the source code, trying to understand the algorithm and advice on the design of the program. We would like to thank Editage (www.editage.jp) for English language editing.